\documentclass[a4paper,12pt]{article}

\usepackage{graphicx}
\usepackage{amsmath,amssymb,bm}
\usepackage{amsfonts}
\usepackage{epstopdf}

\newcommand{\be}{\begin{equation}}
\newcommand{\ee}{\end{equation}}

\newcommand{\ba}{\begin{eqnarray}}
\newcommand{\ea}{\end{eqnarray}}

\begin{document}

\begin{center}
{\large\bf Gluonic Distribution in the Constituent Quark and Nucleon Induced by the Instantons}

{Baiyang Zhang$^{1,2,\P}$}, {Nikolai Kochelev$^{2}$}, {Hee-Jung Lee,$^{3}$}, { Pengming Zhang$^{4}$}

$^{1,4}${Institute of Modern Physics, Chinese Academy of Science, 509 Nanchang Road, 730000, Lanzhou, China}

$^2${Bogoliubov Laboratory of Theoretical Physics, Joint Institute for Nuclear Research, Dubna, 141980,Moscow Region,  Russia}

$^3${Department of Physics Education, Chungbuk National University, Chungbuk, 361-763, Cheongju, Korea}

$^\P${E-mail: zhangbaiyang@impcas.ac.cn}
\end{center}

\centerline{\bf Abstract}
Instanton effects can give large contribution to strong interacting processes, especially at the energy scale where perturbative QCD is no longer valid. However instanton contribution to the gluon contribution in constituent quark and nucleon has never been calculated before.
Based on both the constituent quark picture and the instanton model for QCD vacuum, we
calculate the unpolarized and polarized gluon distributions in the constituent quark and in the nucleon for the first time.
We find that the pion field plays an important role in  producing both the unpolarized and the polarized gluon distributions.

\section{Introduction}

Due to the non-perturbative nature of QCD, when dealing with low-energy states of hadrons, such as nucleons, various models has to be adopted, the most important one is the parton model.
The distribution of partons are described by the parton distribution functions (PDFs), one of the cornerstones of the calculation of high energy cross sections. PDFs are fit from experiment data, not calculated from first principles.
Among various PDFs, the gluon distribution function in the proton give the dominant contribution to the cross sections, they are also of great importance in order to understand the so-called proton spin crisis~\cite{Aidala:2012mv}.
\\
The general form of interaction vertex of massive quark with gluon can be written as
  \be
    V_\mu(k_1^2,k_2^2,q^2) = -g_s t^a \left[ {\gamma_\mu F_1(k_1^2,k_2^2,q^2)} + {\frac{\sigma_{\mu\nu}q^\nu}{2M_q} F_2(k_1^2,k_2^2,q^2)}\right],
  \ee
  where $k_{1,2}^2 $ are the virtualities of incoming and outgoing quarks and $q$ is the momentum transferred.
  The {Anomalous Quark Chromomagnetic Moment (AQCM)} is~\cite{Kochelev:1996pv,diakonov,kochelev4}
  \begin{equation}
    \mu_a=F_2(0,0,0)= -\frac{3\pi (M_q\rho_c)^2}{4\alpha_s(\rho_c)}
  \end{equation}
Based on both the constituent quark model for the nucleon and the instanton model for the non-perturbative QCD vacuum~\cite{shuryak,diakonov}, we can calculate the gluon distribution function in the quarks and nucleons.
The instantons are a result of tunneling effects in QCD, they can be considered as strong non-perturbative fluctuations of the gluon fields in vacuum which describes the non-trivial topological structure of the QCD vacuum.
The average size of the instantons $\rho_c\approx 1/3$~fm is much smaller than the confinement size $R_c\approx 1$~fm.
Furthermore, spontaneously chiral symmetry breaking (SCSB) induced by the instantons is partly responsible for the existence of constituent massive quark.
The SCSB also has important effects in the hadronic reactions, such as in the spin-flip effects observed in high energy nucleon reactions~\cite{kochelev3}.
Instanton contribution to such process was suggested in \cite{kochelev1}.
the connection between effective mass of the quark and the instanton vacuum is discussed in \cite{diakonov,kochelev4,Balla:1997hf,diakonov}.
Instanton contribution to the high energy inclusive pion production in the proton-proton collisions was discussed in ref~\cite{Kochelev:2015pha,Kochelev:2015jba}.

\section{Gluon unpolarized and polarized distributions in the constituent quark}

The effective Lagrangian based on the quark-gluon chromomagnetic interaction that preserves the  chiral symmetry \cite{Balla:1997hf,diakonov} reads
\begin{equation}
{\cal L}_I= -i\frac{g_s\mu_a}{4M_q}\bar q\sigma^{\mu\nu}t^a
e^{i\gamma_5\vec{\tau}\cdot\vec{\phi}_\pi/F_\pi}q G^{a}_{\mu\nu},
\label{Lag2}
\end{equation}
where $\mu_a$ is the (AQCM), $g_s$ is the strong coupling constant, $G^{a}_{\mu\nu}$ is the gluon field strength, and $F_\pi = 93~\mbox{MeV}$ is the pion decay constant.
Expand above Lagrangian to the first order of the pion field, we get~\cite{Kochelev:2015pha,Kochelev:2015jba}
\begin{equation}
\mathcal{L}_I =- i\frac{g_s\mu_a}{4M_q} \, \bar q\sigma^{\mu\nu}t^a q \, G^{a}_{\mu\nu}
+\frac{g_s\mu_a}{4M_qF_\pi}\bar q \sigma^{\mu\nu} t^a \gamma_5
\bm{\tau}\cdot \bm{\pi} q \, G^{a}_{\mu\nu}.
\label{Lag}
\end{equation}
Where $M_q$ is the effective quark mass in the instanton vacuum\cite{Faccioli:2001ug,shuryak}. The strong coupling constant is fixed at instanton scale as
$\alpha_s(\rho_c)=g_s^2(\rho_c)/4\pi \approx 0.5$ \cite{diakonov}.

The Altarelli-Parisi (AP) approach is adopted to calculate the gluon distribution functions in the constituent quark~\cite{Altarelli:1977zs}. The contributions from the perturbative QCD (pQCD) and the non-perturbative instanton-induced interactions are shown in Fig.1.
\begin{figure}[htbp]
\centering
\includegraphics[width=\linewidth]{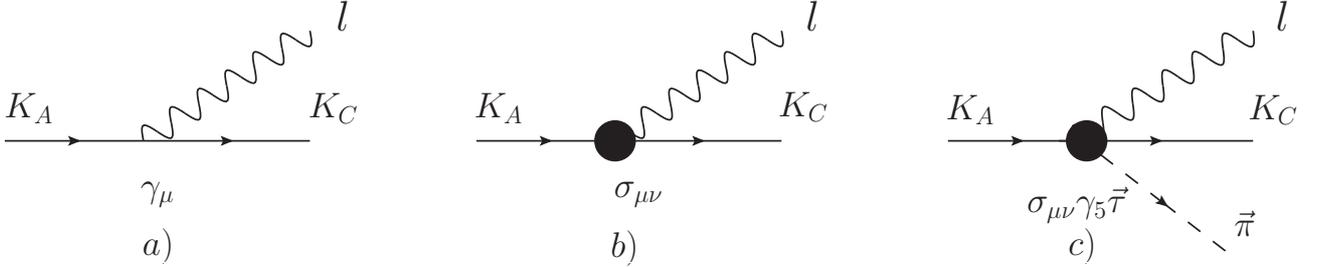}
\caption{a) corresponds to the contributions from the pQCD to gluon distribution in the quark.
b) and c) correspond to the contributions to the gluon distribution in the quark
from the non-perturbative quark-gluon interaction and the non-perturbative quark-gluon-pion interaction, respectively.}
\label{fig:1}
\end{figure}
In the single instanton approximation, the expansion parameter is $\delta = (M_q\rho_c)^2 $, which is relatively small with the value of the quark mass $M_q=86 $ MeV in the effective single instanton approximation which we used here~\cite{Faccioli:2001ug}.

The non-perturbative contribution with the pion, i.e. diagram c) in Fig.1, has never been calculated before.
By calculating the matrix element for the diagram c) in Fig.1 and performing the integration over $x$, we can obtain the unintegrated unpolarized and polarized gluon distribution in the quark.

\begin{figure}[h]
\begin{minipage}[c]{0.33\linewidth}
\centering{\includegraphics[width=5cm,height=3cm,angle=0]{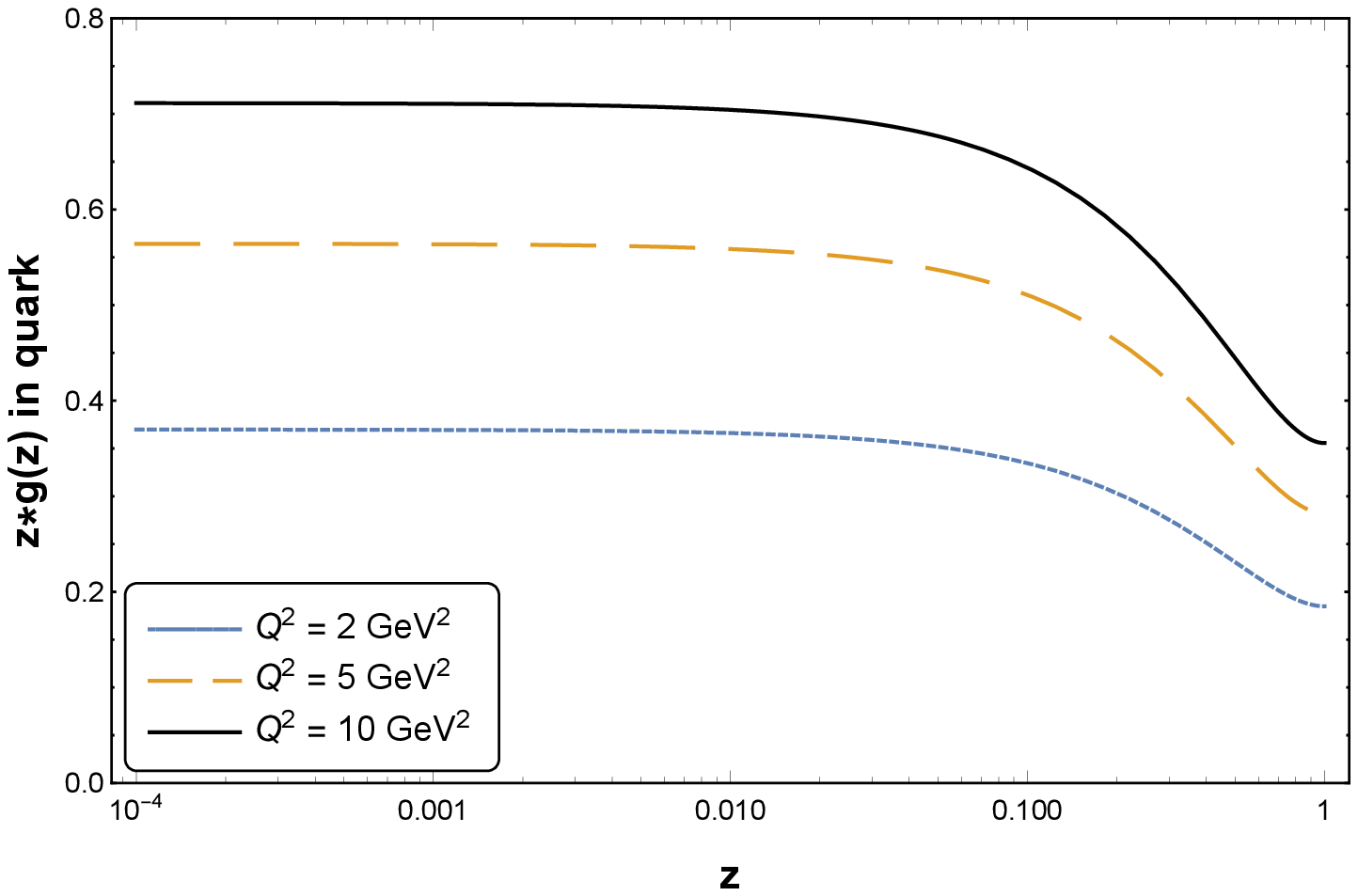}}\
\end{minipage}%
\begin{minipage}[c]{0.33\linewidth}
\centering{\includegraphics[width=5cm,height=3cm,angle=0]{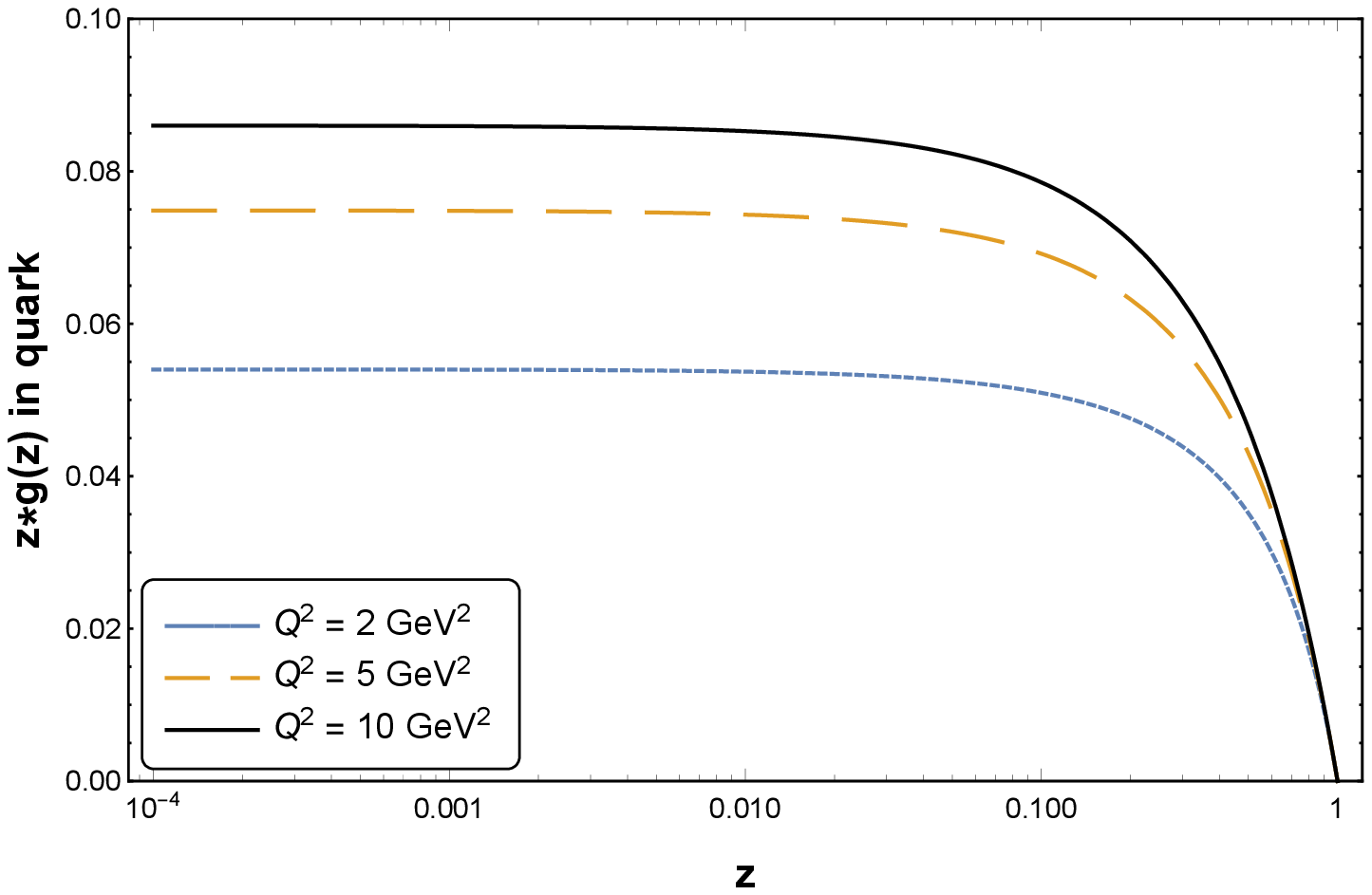}}\
\end{minipage}%
\begin{minipage}[c]{0.33\linewidth}
\centering{\includegraphics[width=5cm,height=3cm,angle=0]{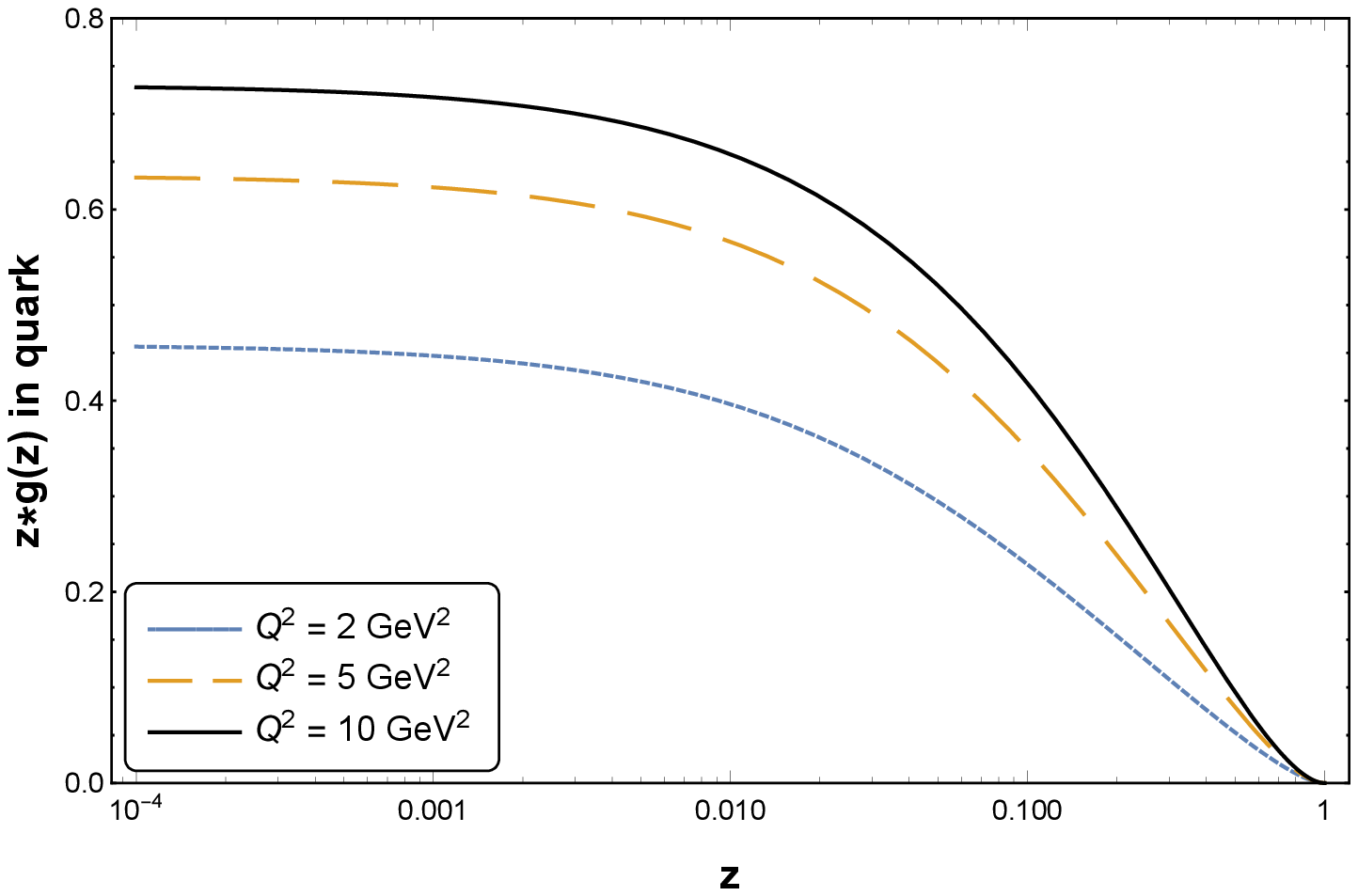}}\
\end{minipage}
\caption{ The $z$ dependency of the contributions to the the unpolarized gluon distribution
in the constituent quark from the pQCD (left panel),
from the non-perturbative interaction without pion (central panel), and from the non-perturbative
interaction with pion (right panel). The dotted line corresponds to $Q^2=2$ GeV$^2$,
the dashed line to $Q^2=5$ GeV$^2$, and the solid line to $Q^2=10$ GeV$^2$.}
\end{figure}

\begin{figure}[h]
\begin{minipage}[c]{0.5\linewidth}
\centering{\includegraphics[width=\linewidth]{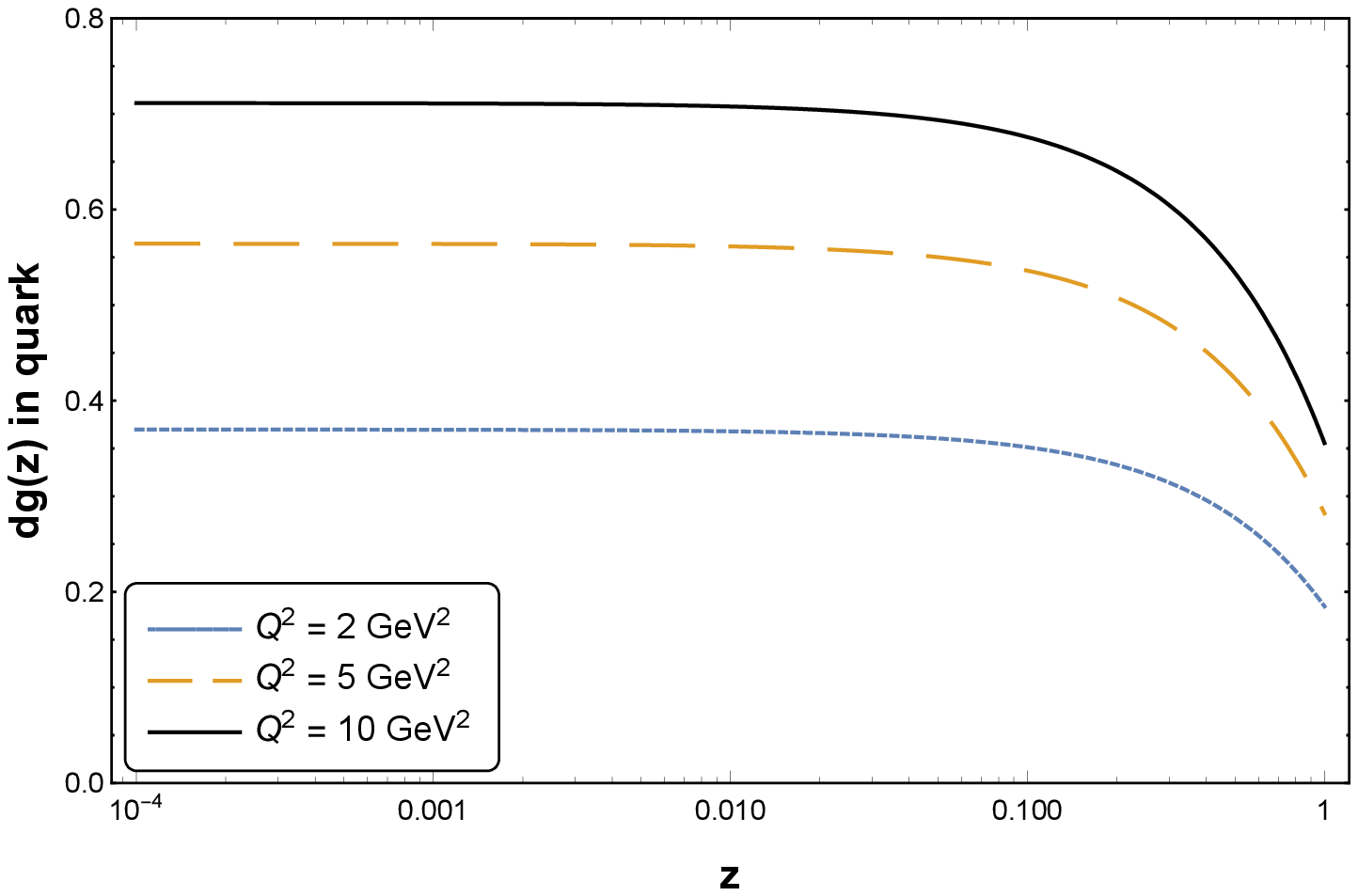}}\
\end{minipage}%
\begin{minipage}[c]{0.5\linewidth}
\centering{\includegraphics[width=\linewidth]{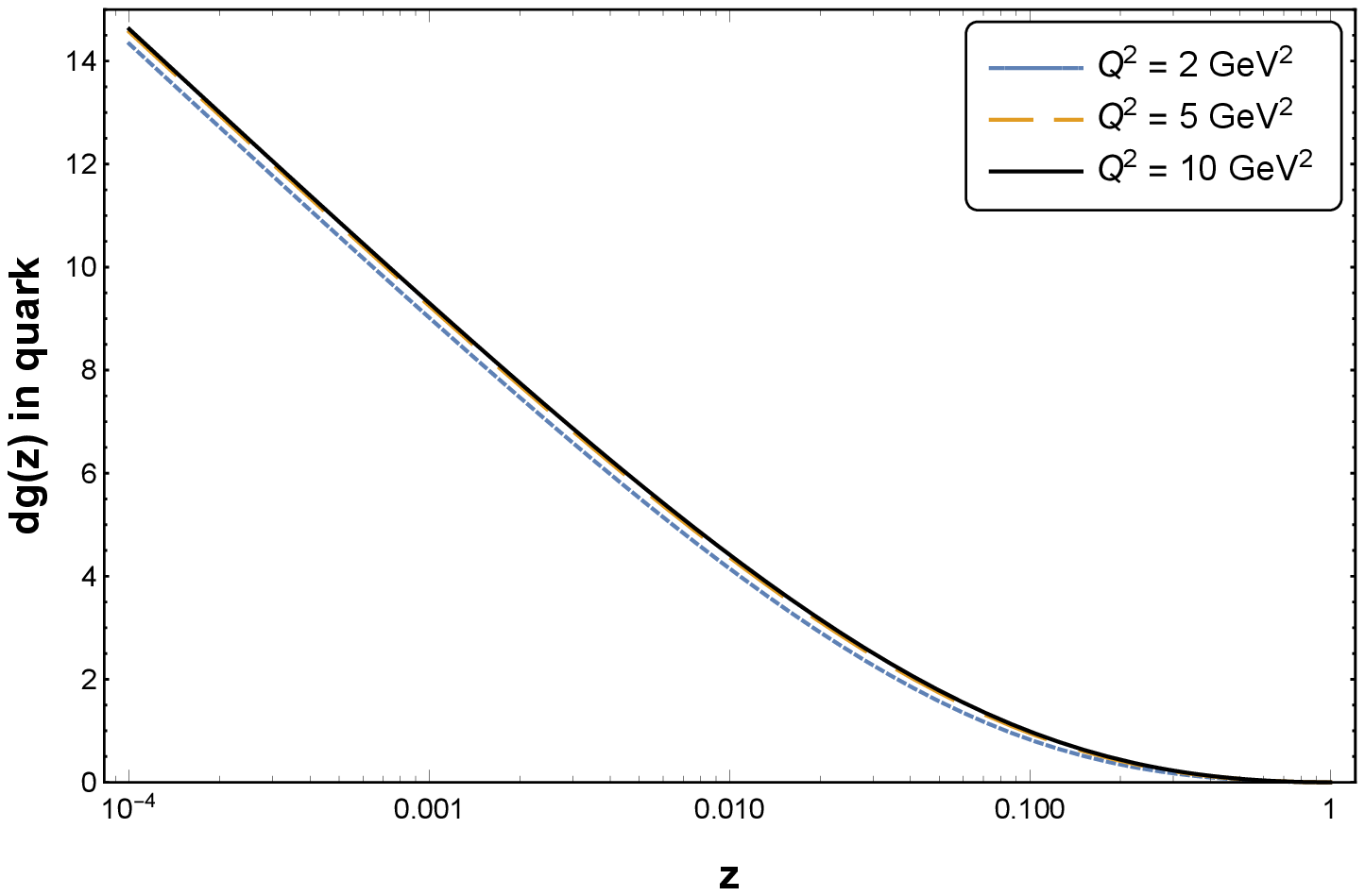}}\
\end{minipage}
\caption{ The $z$ dependency of the contributions to the polarized gluon distribution
in the constituent quark from the pQCD (left panel), and from the non-perturbative
interaction with pion (right panel). The notations are the same as in the Fig.2. In the right panel the result for
$Q^2=5 $ GeV$^2$ does not shown because it is practically identical to the $Q^2=10 $ GeV$^2$ case. }
\end{figure}

Fig.2 shows the unpolarized gluon distribution in constituent quark versus $z$ at several $Q^2$, where $z$ is the fraction of quark momentum carried by gluon.

On one hand, the unpolarized gluon distribution produced by the perturbative QCD depends stronger on $z$ than that produced by the non-perturbative interactions.
Hence the pQCD dominates the large $z$ region.
On the other hand, the gluon distribution produced by non-perturbative quark-quark-gluon interaction is rather small in comparison with that from both the pQCD, a) in Fig.1, and the non-perturbative interaction with pion, c) in Fig.1, partly because of a larger final phase space due to an extra pion.
At small $z$ for all the contributions $zg(z,Q^2)\approx const$, which is so-called Pomeron-like behavior.
The polarized gluon distributions in the constituent quark are shown in Fig.3.
Again it is obvious that pQCD contribution to the polarized gluon distribution dominates
in the large $z$ region, where the non-perturbative interaction with pion dominates in small $z$ region.
Furthermore, pQCD contribution shows $\Delta g(z,Q^2)\rightarrow  const$,
while the non-perturbative contribution shows anomalous dependency
as $\Delta g(z,Q^2)\rightarrow \log(z)$.
When $Q^2$ increases, the pQCD contribution increases as $\log(Q^2\rho_c^2)$ while the non-perturbative contribution practically stays the same for $Q^2>1/\rho_c^2=0.35$GeV~$^2$.
Therefore, the non-perturbative interaction contribution to the gluon distribution can be ragarded as an intrinsic polarized gluon inside the constituent quark.

\section{Gluon distributions in the nucleon}
Knowing the gluon distribution in the constituent quark, we can use convolution model to obtain the gluon distribution in the nucleon.
The PDF of the unpolarized constituent quark is taken to be
\be
 q_V(y)=60y(1-y)^3.
\label{unpolvalence}
\ee
This PDF is in accord with the quark counting rule at large $y$ this distribution.
Its low $y$ behavior and normalization are fixed by the requirements that $\int_0^1dyq_V(y)=3$ and $\int_0^1dyyq_V(y)=1$.
For the polarized constituent quark distribution, a simple form is adopted to be
\be
 \Delta q_V(y)=2.4(1-y)^3.
\label{polvalence}
\ee
Which is also in agreement with the quark counting rule at $y\rightarrow 1$.
The normalization conditions has been fixed from the hyperon weak decay data (see \cite{Aidala:2012mv}) as
\be
\int_0^1dy\Delta q_V(y)=\Delta u_V+\Delta d_V\approx 0.6.
\label{initialpol}
\ee

\begin{figure}[h]
\begin{minipage}[c]{0.5\linewidth}
\centering{\includegraphics[width=\linewidth]{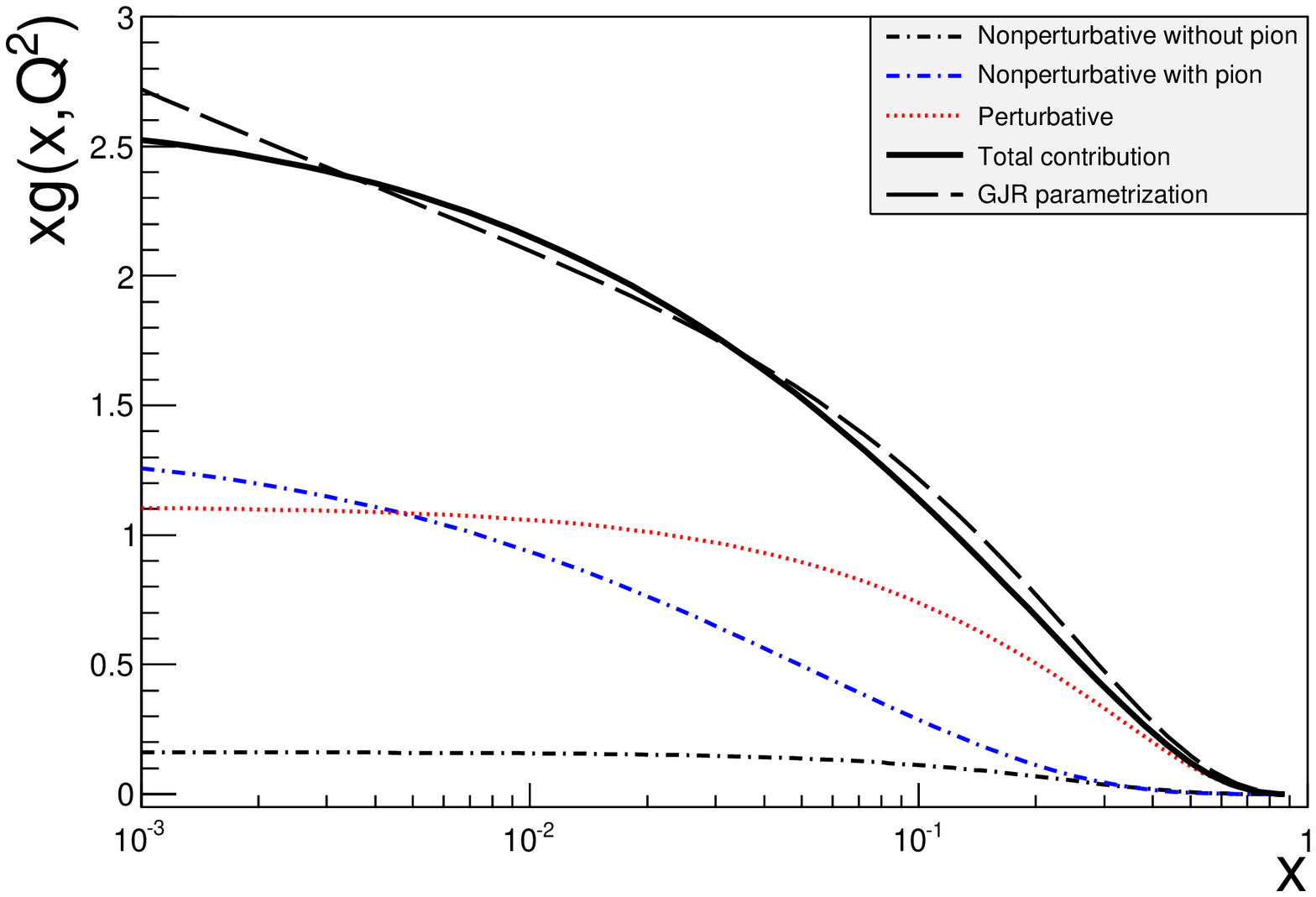}}\
\end{minipage}%
\begin{minipage}[c]{0.5\linewidth}
\centering{\includegraphics[width=\linewidth]{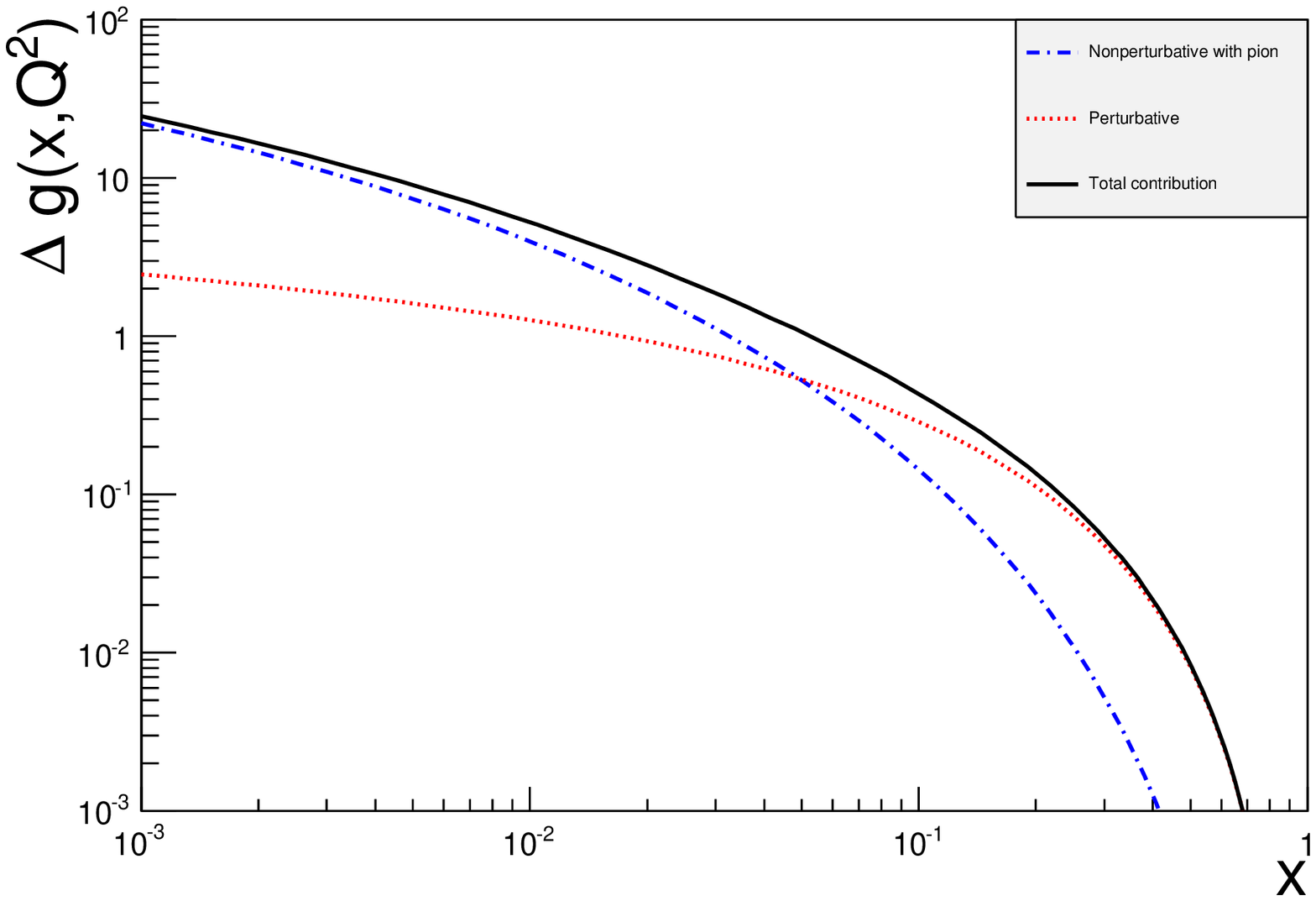}}\
\end{minipage}
\caption{ Unpolarized (left panel), and the  polarized
(right panel) gluon distributions in the nucleon at the scale $Q^2=2$ GeV$^2$. The dotted line in red corresponds to
the contribution from pQCD, the dotted-dashed in blue to the contribution from the non-perturbative interaction with pion,
dotted-dashed in black to the contribution from the non-perturbative interaction without pion,
and the solid line to the total contribution.}
\end{figure}

Both the unpolarized and polarized gluon distributions in the nucleon have been presented in Fig.4, where $Q_0^2=2$ GeV$^2$. This value of $Q^2$ is often used as the starting point for the standard pQCD evolution.
Our result for the unpolarized gluon distribution, which is shown in the left panel of Fig.4, is identical to the GJR parametrization $ xg(x)=1.37x^{-0.1}(1-x)^{3.33} $~\cite{Gluck:2007ck}.
It is well known that the behavior of non-polarized gluon distribution at low $x$ region is determined
by the exchange of Pomerons.
Pomeron exchange effects play a very important role in the phenomenology of high energy reactions.
Our results shown in Figs.2,4 indicate the existence of two different kinds of Pomerons, the "hard" pQCD Pomeron
and the "soft" non-perturbative Pomeron, they have quite different dependency on $x$ and $Q^2$.
Our calculation strongly supports two kinds of Pomerons pictures which can explain many experimental data in both the DIS data at large $Q^2$ and the high energy cross sections with small momentum transfer~\cite{Landshoff:2009wt}.

\section{Proton spin problem}

The polarized gluon distribution is shown in the right panel of Fig.4.
The fraction of nucleon momentum that carried by the gluons $G(Q^2)=\int_0^1dxxg(x,Q^2)$, as well as their polarization $\Delta G(Q^2)=\int_0^1dx \Delta g(x,Q^2)$ are presented as a function of $Q^2$ in Fig.5.
It is clear that the non-perturbative interaction without pion gives a small contribution to
unpolarized gluon distribution.
Furthermore, such contribution is zero for the polarized gluon case.
For $Q^2>1/\rho_c^2$, the main contributions to both the unpolarized and the polarized gluon distributions come from pQCD and the non-perturbative interaction with the pion.

\begin{figure}[h]
\begin{minipage}[c]{0.5\linewidth}
\centering{\includegraphics[width=\linewidth]{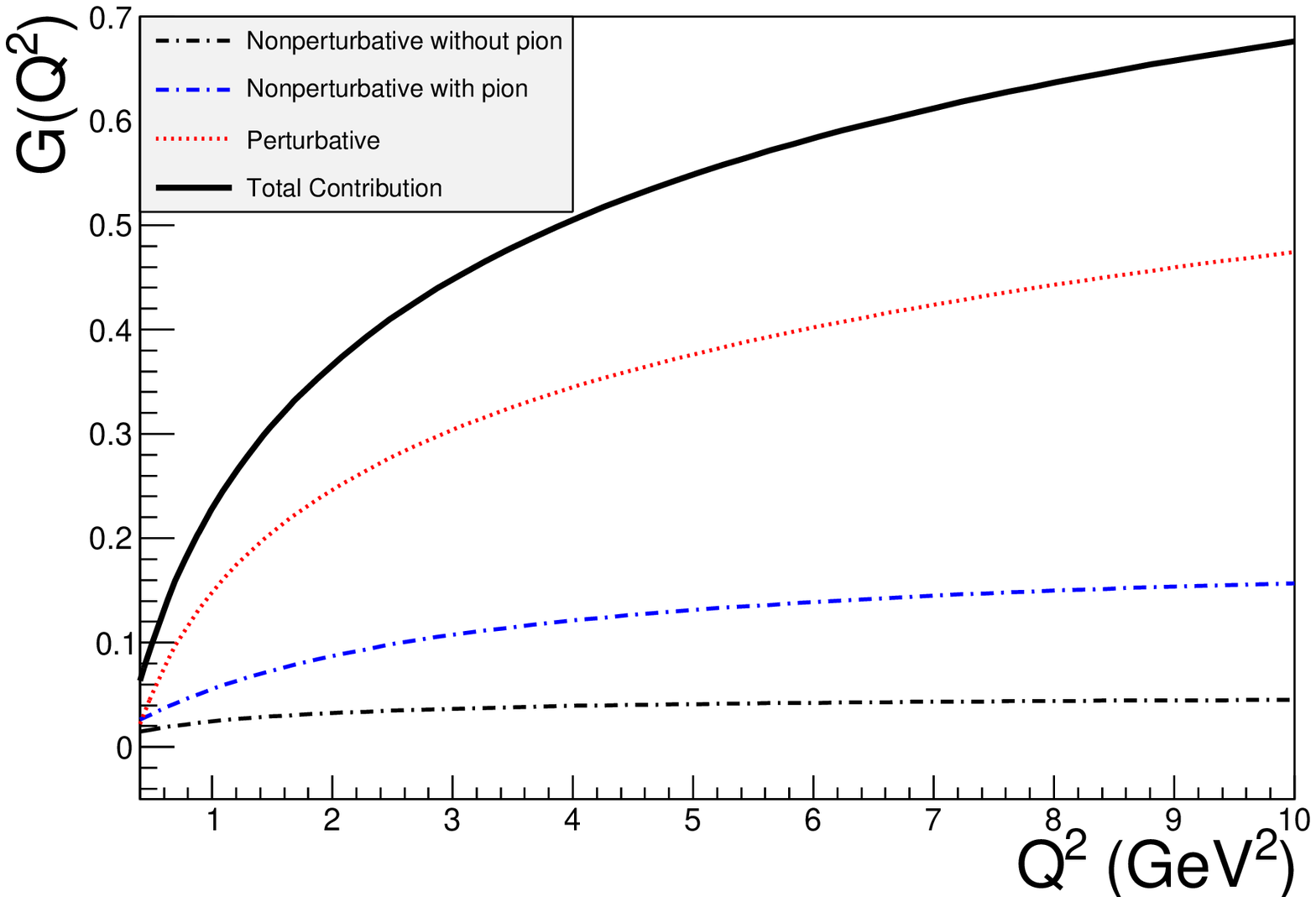}}\
\end{minipage}%
\begin{minipage}[c]{0.5\linewidth}
\centering{\includegraphics[width=\linewidth]{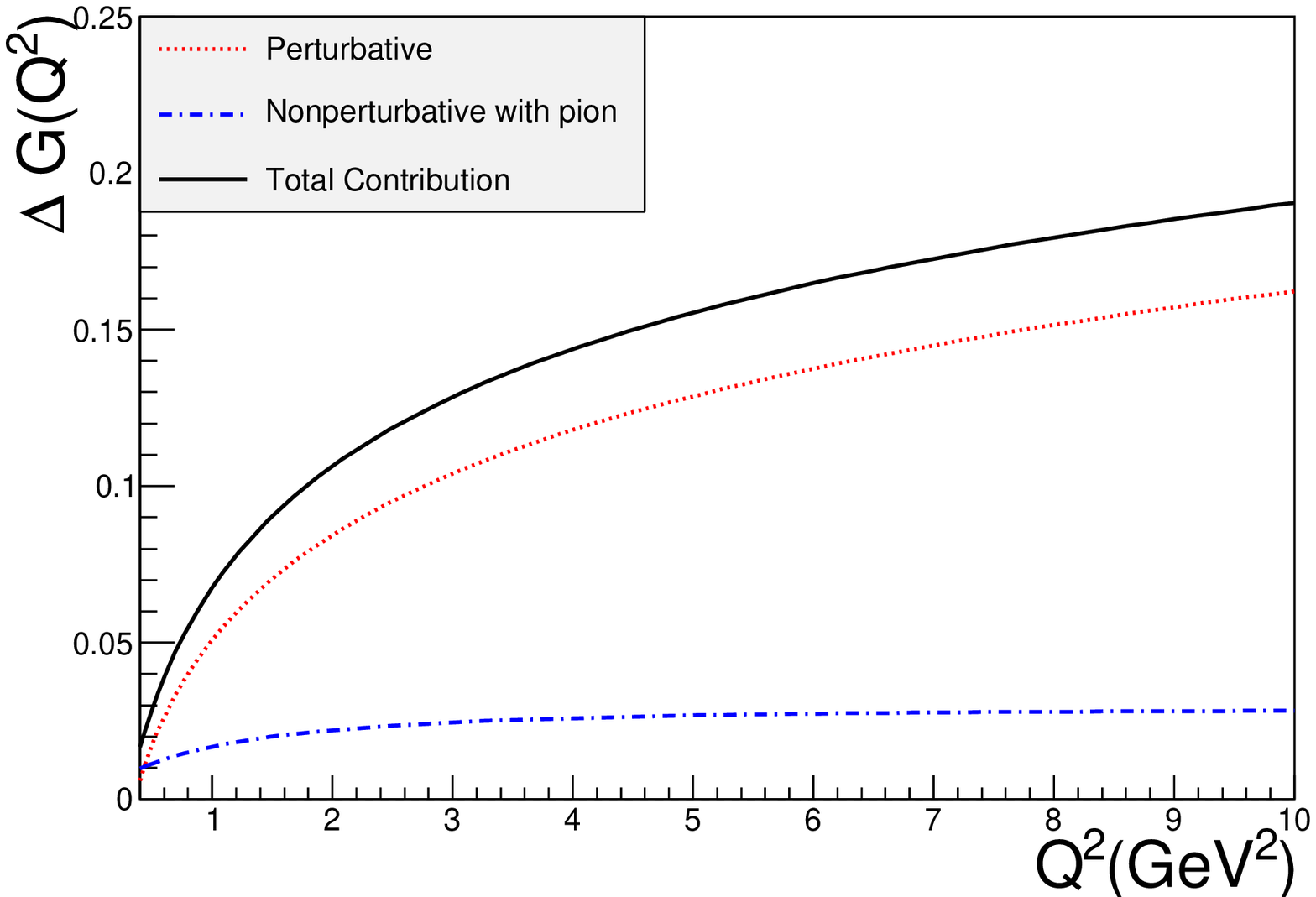}}\
\end{minipage}
\caption{ The part of the nucleon momentum carried by gluons (left panel), and the contribution of the gluons
to nucleon spin (right panel) as the function of $Q^2$.
The notations are the same as in the Fig.4 }
\end{figure}

Jaffe and Manohar decomposed the proton spin into\cite{Jaffe:1989jz}
\be
\frac{1}{2}=\frac{1}{2}\Delta\Sigma+\Delta G+L_q+ L_g,
\label{spin}
\ee
where the first term is quark spin, $\Delta G=\int_0^1dx\Delta g(x)$ is the gluon polarization and the last two terms are quark and gluon orbital momentum.
The key problem is to explain discovered small value of the proton spin carried by quark.
At present we have $\Delta\Sigma \approx 0.25$ \cite{Aidala:2012mv}, which is in bad agreement with $\Delta\Sigma=1$ given by the non-relativistic quark model.
Shortly after the EMC data which reveals such discrepancy between experiment and theoretical predictions, the axial anomaly effect in DIS was brought out and considered to be the primary candidate as a remedy for the dilemma. \cite{Altarelli:1988nr}.
For three quark flavors, it gives following reduction of the quark helicity in the DIS
\be
\Delta\Sigma_{DIS}=\Delta\Sigma-\frac{3\alpha_s}{2\pi}\Delta G
\label{DIS}
\ee
A big positive gluon polarization is needed, $\Delta G\approx 3\div 4$, in order to explain the small value of $\Delta\Sigma_{DIS}$.
However, modern experimental data from the inclusive hadron productions and the jet productions have exclude such a
large gluon polarization in the accessible intervals of $x$ and $Q^2$ \cite{deFlorian:2014yva,Nocera:2014gqa}, and so does our model (see Fig.5, right panel).
Therefore, the axial anomaly effect, suggested in \cite{Altarelli:1988nr}, cannot explain the proton spin problem.
We should stress that the helicity of the initial quark is flipped in the vertices b) and c) in Fig.1.
As the result, such vertices should lead to the {\it screening} of the quark helicity. It is evident that at the $Q^2\rightarrow 0$ such screening vanishes as shown in Fig.5 and the total spin of the proton  is carried by its constituent quarks.

We would like to emphasis that in our approach, our model enables us to calculate the un-integrated gluon distribution function, which is of essential importance in many application, such as the calculations of high energy various reactions.
\section{Conclusion}
We show that the quark-gluon-pion anomalous interaction gives a very large contribution to  both
the unpolarized and polarized gluon distributions. It means that pion field plays a fundamental
role to produce  both gluon distributions in hadrons.
The possibility of the matching of the constituent quark model for the nucleon with
its partonic picture is shown. The phenomenological arguments in favor of such a non-perturbative
gluon structure of the constituent quark were recently given in the papers \cite{Kopeliovich:2007pq,Kopeliovich:2006bm}.
We also pointed out that the famous proton spin crisis might be explained by the
flipping of the helicity of the quark induced by non-perturbative anomalous quark-gluon and quark-gluon-pion interactions.

\section*{Acknowledgments}
We are grateful to Igor Cherednikov, Boris Kopeliovich, Victor Kim and Aleksander Dorokhov for useful discussions.
 This work was partially supported by the National Natural
  Science Foundation of China (Grant No. 11575254 and 11175215), and by the Chinese Academy of Sciences visiting
professorship for senior international scientists (Grant No.
2013T2J0011). This research was also supported in part by the
Basic Science Research Program through the National Research
Foundation of Korea(NRF) funded by the Ministry of
Education(2013R1A1A2009695)(HJL).

\end{document}